\documentstyle[12pt]{article}
\def\b{\beta}
\def\g{\gamma}
\def\d{\delta}
\def\h{\eta}
\def\q{\theta}
\def\bq{\bar{\theta}}
\def\y{\psi}
\def\by{\bar{\psi}}
\def\r{\rho}
\def\p{\pi}
\def\f{\phi}
\def\m{\mu}

\def\bg{\begin{eqnarray}}
\def\ed{\end{eqnarray}}

\begin{document}
\begin{flushright}
SSCL\\
SFU-Preprint-92-5 \\
\end{flushright}
\vskip.11in
\begin{center}
{\Large\bf Comments on the Vacuum Orientations in QCD}
\\*[.25IN]
{\bf Zheng Huang} and {\bf K.S. Viswanathan}\\
Department of Physics, Simon Fraser University\\
 Burnaby, B.C. V5A 1S6, Canada\\
\vskip .2in
{\bf Dan-di Wu}$^*$\\
Superconducting Super Collider Laboratory$^\dagger$\\
2550 Beckleymeade Avenue, Dallas, Texas 75237\\
\end{center}
\vskip .35in

\begin{abstract}
We study the QCD vacuum orientation angles in correlation with the strong
CP phases. A vacuum alignment equation of the dynamical chiral symmetry
breaking is derived based on the anomalous Ward identity.
It is emphasized that a chiral rotation of the quark field
causes a change of the vacuum orientation and a change in the definition
of the light pseudoscalar generators. As an illustration of the idea,
 $\h\rightarrow 2\p$ decays are carefully studied in different chiral frames.
Contrary to the claim in Ref.[7], the $\theta$-term does not directly
contribute to the CP-violating processes.
\end{abstract}
\vskip .5in
\begin{flushleft}
\footnotesize{\today}\\
\end{flushleft}
\begin{small}\rule{1.5in}{.01in}\\
$^*$ \footnotesize{On leave of absence from School of Physics, University of
Melbourne,
Parkville, Vic 3052, Australia.} \\
$^\dagger$ \footnotesize{Operated by the University Research Association,
In., for the U.S. Department of Energy under Contract No. DE-AC35-89ER40486.}
\end{small}
\clearpage
In QCD lagrangian of strong interactions, there are two possible
sources of CP violation: the
complex quark mass terms and the $\q$-term. It has been long realized that they
are related to each other by  chiral transformations associated with the quark
fields. The physical effects of  CP violation only
depend on a chiral-rotation
invariant $\bq$ defined as
\begin{equation}
\bq = \q_{QCD} +\q_{QFD} =\q_{QCD} +\sum_{i}^{L_f}\phi_{i}
\end{equation}
where $\q_{QCD}$ is the coefficient of the $\q$-term, $\phi_{i}$ is the phase
of
the ith quark mass term, and $L_f$ is the number of light quarks\footnote{The
inclusion of heavy quarks will not change our discussion significantly if they
are in the normal phase. Otherwise see Ref.\ [4].}.
However, there is another source of CP-violating angles, the phases of the
quark condensates, which arise from  dynamical chiral symmetry breaking
(DCSB)
\begin{equation}
\langle \by_{L}^{i}\y_{R}^{i}\rangle =-\frac{C_{i}}{2}e^{i\b_i}\; ; \;
\langle \by_{R}^{i}\y_{L}^{i}\rangle =-\frac{C_{i}}{2}e^{-i\b_i}
\end{equation}
where $\y$ is the quark field and $C_i$'s, $\b_i$'s are real. The QCD vacuum
orientation is characterized by a set of phases of the quark condensates. If
the vacuum angle $\b_i \neq 0$
it follows that $\langle \by^{i}i\g_5\y^{i}\rangle =
C_{i}\sin{\b_i}\neq 0$ which may also
break CP symmetry since $\langle \by i\g_5\y\rangle$ is a P-odd and CP-odd
quantity.
It has been proven by Vafa and Witten that when $\bq= 0$ and $\f_i=0$ for all
$i$'s, the parity
symmetry in a vector-like theory such as QCD is not spontaneously broken
\cite{witten} therefore $\b_i=0$ or $\pi$ \cite{sikivie} for all $i$'s.
When $\bq\neq 0$, on the other hand, one generally expects that the
CP-violating interactions
in the lagrangian may result in a CP-asymmetric physical vacuum. The purpose of
the present paper
is to study the vacuum orientation in the presence of strong CP violation and
its
potential effects on CP-violating processes in strong interactions.
We find that the phases of quark condensates can be completely determined
as functions of $\bq$ and $\f_i$'s via a vacuum alignment equation.
 Thus $\b_i$'s are not spontaneously generated either even when
the CP symmetry is explicitly violated by $\bq\neq 0$.

Obviously, the quark condensates (2) cannot be referred
to as fundamental  parameters of the theory
since they are subject to  chiral transformations.
 In fact $\b_i$'s can be set to any values if we make appropriate
chiral transformations for the quark fields.
Such transformations also change the phases of the quark masses, as well as
the coefficient of the $\q$-term because of the chiral anomaly. But what is
important is
the correlation between the vacuum orientation and the distribution of
the strong CP phases among $\q$-term and quark mass terms, which is to be
determined by the vacuum alignment. The effective CP-violating interactions
in low energy hadron physics (for instance, in current algebra) highly depend
on this correlating feature. As we shall see below,  the sole $\bq$-dependence
of the
strong CP effects is proven only when  the orientations of the vacuum are
properly
considered. In addition, it is of interest to study  DCSB in the presence
of strong CP violation in its own right.

One way  of relating the phases of the quark condensates with
$\q_{QCD}$ and $\f_i$'s
is to consider the so-called anomalous Ward identity \cite{1}
\begin{eqnarray}
\frac{1}{2}\partial^\m J^{(i)5}_\m &=&F\tilde{F}+
im_ie^{i\f_i}\by_{L}^{i}\y_{R}^{i}
-im_ie^{-i\f_i}\by_{R}^{i}\y_{L}^{i}\\
 & &(i=1, 2, \ldots, L_f) \nonumber
\end{eqnarray}
where $J^{(i)5}_\m=\by^i\g_\m\g_5\y^i$,
$F\tilde{F}=\frac{g^2}{32\p^{2}}\epsilon_{\m\nu\r\sigma}F^{\m\nu}F^{\r\sigma}$
and $F^{\m\nu}$ the non-abelian gauge field strength tensor.
 Taking the vacuum
expectation values (VEV) on both sides of (3) yields \cite{2}
$$\left\langle {F\tilde F} \right\rangle=-im_i[e^{i
\phi _i}\left\langle {\bar \psi _L^i\psi _R^i}
\right\rangle-e^{-i\phi _i}\left\langle {\bar \psi _R^i\psi _L^i}
\right\rangle] \eqno(4a)$$
$$\quad\quad\quad\quad\quad\quad=-m_iC_i\sin (\phi _i+\beta
_i)\quad\quad\,\,(i=1,2,\cdots ,L_f). \eqno(4b)$$
In deriving (4a) we have assumed that the VEV's of the divergence
of the gauge invariant
current vanish. Eq.(4b) is the master equation of this paper.
It is important to point out that if
the DCSB does not occur, (4a) would vanish identically and there is no
constraint on those phases.
Indeed even though the quark condensate can be non-zero
due to the explicit chiral symmetry
breaking (ECSB) i.\ e.\  the quark current masses,  it does not
contribute to (4a) because it
possesses a phase opposite
to the phase of the quark mass $\f_i$ and renders the RHS of (4a) zero.
This can be easily seen by taking the free-quark limit in which the condensate
is calculated as
$$\left\langle {\bar \psi _L^i\psi _R^i} \right\rangle=
{\textstyle{1 \over 2}}\int {{\textstyle{{d^4k}
\over {(2\pi )^4}}}}Tr{{1+\gamma _5} \over
{\not k-m_ie^{i\phi _i\gamma _5}}}=-m_i\Lambda ^2(m_i)e^{-i\phi _i}
\eqno(5)$$
where $\Lambda^2$ is real. The substitution of (5) into (4a) yields $\langle
F\tilde{F}\rangle=0$. Therefore, $C_i$'s
 in (4a) should be understood
as the purely $dynamical$ condensates
 originating from DCSB.
The $kinematical$ part of the condensates that is induced
by the ECSB and has a phase $-\f_i$ has been subtracted out in (4a). It is  the
DSCB
combining with the topological structure of QCD
Yang-Mills fields (the instanton effect) that makes
the strong CP phases non-trivial and relates them to each other.

Eq.(4b) is not immediately useful to us since it has an unknown quantity
$\langle F\tilde{F}\rangle$. It involves no quark fields thus is independent
of the chiral transformation. It is conceivable that $\langle
F\tilde{F}\rangle$
is  solely a function of $\bq$ (not of $\f_i$'s
and $\b_i$'s separately). A rigorous proof
can be made by summing over instanton configurations
in QCD $\q$ vacua. For simplicity,
consider QCD with a single quark field $\y$. The VEV's of $F\tilde{F}$ is
given by \cite{3}
$$\left\langle {F\tilde F} \right\rangle={1 \over {VT}}\left\langle
{\int {d^4}xF\tilde F} \right\rangle\quad\quad\quad\quad
\quad\quad\quad\quad\quad\quad\quad\quad\quad\quad\quad\quad\quad\quad\eqno(6)$$
$$\quad\quad\quad\quad\;={1 \over {VT}}{1 \over N}\sum\limits_{\nu =0,\pm
1,\cdots }
{e^{i\bar \theta \nu }\nu \int {[dA_\mu ]_\nu \det (i\not\!\!
 D_\nu +im) \exp(-\int {d^4xFF)}}}$$
where $N$ is the normalization factor, $VT$ is the volume of
Euclidean space-time,
and $\nu$ is the winding number of the instanton field configuration,
and the fermion determinant results
from the integration over the quark field. We have made an appropriate chiral
transformation such that the quark mass is real and $\q_{QCD}=\bq$ (we
can always do so because the generating functional is invariant under the
redefinition of integral variables). It is shown that
when $\nu>0$ ($<0$) $i\not\!\! D_\nu$ has $|\nu|$ zero modes with
negative (positive)
chirality \cite{4}. We  thus obtain
$$\left\langle {F\tilde F} \right\rangle=mA_1(m^2){\textstyle{{-i}
\over 2}}(e^{i\bar \theta }-e^{-i\bar \theta })\quad\quad\quad\quad\quad\quad
\quad\quad\quad\quad\quad\quad\quad\quad\quad\quad\quad\quad$$
$$\quad\quad\quad+m^2A_2(m^2){\textstyle{{-i} \over 2}}(e^{i2\bar \theta
}-e^{-i2\bar
\theta })+\cdots +m^{|\nu|}A_{|\nu|} (m^2){\textstyle{{-i} \over 2}}
(e^{i|\nu| \bar \theta }-e^{-i|\nu| \bar \theta })+\cdots $$
$$\quad\quad\;=mA_1(m^2)\sin \bar \theta +m^2 A_2(m^2) \sin{2\bar{\theta}} +
\cdots =K(m,\q)\sin\bar \theta \quad\quad\quad\quad\eqno{(7)}$$
where $A_{|\nu|}(m^2)$'s are given in Euclidean space
$$A_{|\nu|} (m^2)={\textstyle{1 \over {VTN}}}e^{-|\nu| {\textstyle{{8\pi ^2}
\over {g^2}}}}\int {[dA_\mu ]_\nu \prod\limits_{\lambda _r>0}
{[\lambda _r^2(A)+m^2]\exp [-\int {d^4xFF]}}}.\eqno(8)$$
Here $\lambda_r(A)$'s are non-zero eigenvalues of $i\not\!\! D_\nu$. Clearly
$A_{|\nu|}(m^2)$'s are some real functions of $m^2$ and do not vanish as
$m\rightarrow 0$. If $\bq$ is small as it must be, $\langle
F\tilde{F}\rangle\simeq
K(m)\bq=K(m)\q_{QCD}+K(m)\q_{QFD}$.

Combining (7) with (4b), we derive the so-called vacuum alignment equation
(VAE) \cite{2,6}, which determines the orientation of
the QCD vacuum in the presence of strong CP violation
$$K(m)\bar
\theta =m_iC_i(\phi _i+\beta _i)+O(m^2;\bar \theta ^2)\eqno{(9)}$$
$$(i=1,\;2,\;\cdots ,\;L_f).$$
Eq.(9) has proven that $\b_i$'s are not spontaneously
generated even when $\bq\neq 0$. The conclusion of Vafa and
Witten's theorem \cite{witten} can be extended to the case where parity
symmetry is explicitly violated.
A similar result has been worked out previously \cite{6}
from different points of view.
If $m_i$'s vanish, $\b_i$'s can be arbitrary. This is referred to as the
degeneracy of QCD vacua when the ECSB is absent. Any vacuum characterized by a
set of the vacuum angles $\b_i$'s is as good as others and the orientation of
the DCSB is arbitrary. However, the importance of (9) is
that when the ECSB is turned on, the ground
state must align with it in such a way that (9) is satisfied. Though
both $\f_i$ and $\b_i$ are not physical parameters and can be changed through
chiral rotations, their  sum  is uniquely determined by the physical
parameter $\bq$.
When one is chosen the other is completely determined through making the
vacuum alignment.
As is emphasized by Dashen \cite{7}, a misaligned vacuum, whose
 orientation angles do not satisfy (9), may cause an
inconsistency such as the goldstone bosons (pions) acquiring negative mass
squares.

Once the DCSB and the ECSB align with each other, an absolute rotation of the
whole system is of no concern. Thus a chiral transformation is allowed only
if the corresponding change of the vacuum orientation has
been taken into account.
We can have two ways to make the vacuum alignment. We may choose one particular
vacuum, for example, by requiring the quark condensate to be real $\b_i=0$
$(i=1,2,\cdots,
L_f)$ and ask what perturbation (the ECSB) is aligned with it. Recalling that
$C_i$'s are dynamical condensates and thus
$C_i=C_j=C$ $(i,j=1,\;2,\;\cdots ,\;L_f)$,
we obtain by solving (9) for $\f_i$'s,  to $O(m;\bq)$
$$(A)\quad\quad\quad\;\phi _i=
{K(m) \over {m_i}}{\bar \theta }\;\;;\;\;\beta _i=0
\quad\quad(i=1,\;2,\;\cdots ,\;L_f)$$
$$\theta _{QFD}=\sum\limits_i {\phi _i}=\frac{K(m)}{\bar m}\bar
\theta \;\quad;\;\quad\theta _{QCD}= \bar \theta- \theta _{QFD}=\frac{K(m)-\bar
m}
{\bar m}\bar \theta \eqno(10)$$
where $\bar m=(\sum\limits_i {\frac{1}{m_i}})^{-1}$ and the CP-violating
lagrangian
$${\cal L}_{(A)}^{CP}=-\sum\limits_i {m_i\phi _i}\bar
\psi ^ii\gamma _5\psi ^i+\theta _{QCD}F\tilde F=-{K(m)
\over {\bar m}}\bar \theta \,\bar \psi \kern 1pti\gamma _5I\psi
+\frac{K(m)-\bar m}{\bar m}\bar \theta F\tilde F \eqno{(11)}$$
where $I$ is an identity matrix. We shall call the solution (10) basis (A).
Another way is to assume a certain pattern of the ECSB and to ask which one of
the degenerate vacua corresponds to the perturbation. For example, we may
choose
the  quark mass terms real $\f_i=0$ $(i=1,2,\cdots,L_f)$
and determine the vacuum
angle $\b_i$'s. Again, from (9) we have
$$(B)\quad\quad\quad\;\phi _i=0\;\quad;\;\quad\beta _i
={K(m) \over {m_i}}\bar \theta
\quad\,(i=1,\;2,\;\cdots ,\;L_f)$$
$$\theta _{QFD}=\sum\limits_i {\phi _i}=0\;
\quad;\;\quad\theta _{QCD}=\bar \theta \quad\quad\quad\;\eqno(12)$$
and
$${\cal L}_{(B)}^{CP}=\bar \theta \kern 1ptF\tilde F.\eqno{(13)}$$
Solution (12) is to be called basis (B). We would like to emphasize again
that by performing the chiral rotation on quark fields one has switched the
strong
CP phases among the $\q$-term and quark mass terms, and obtained different
lagrangians, each of which corresponds to a certain vacuum orientation. When
calculating the strong CP effects we must take this into consideration to
assure
the correct result.

As an illustration, we compute the CP-violating $\h\rightarrow 2\p$ decays
in two bases with a given $\bq$. In basis (A) where the condensates are real,
we apply the soft-pion theorem to extracting $\h$ and $\p$'s\footnote{We
have worked in the context of $SU(3)_L \times SU(3)_R$ where $\h$ and $\p$'s
are
all light pseudoscalars.}
$${\cal A}(\eta \to \pi ^+\pi ^-)=\left\langle {\pi ^+\pi ^-\left|
{{\cal L}_{(A)}^{CP}} \right|\eta } \right\rangle\quad\quad\quad\quad\quad
\quad\quad\quad\quad\quad\quad\quad\quad\quad\quad\quad\quad\quad\quad\quad$$
$$\;\quad\quad\quad\quad=-{K(m) \over {\bar m}}\bar \theta
({{-i} \over {F_\pi }})^3\left\langle {[Q_5^8,\kern 1pt[Q_5^+,
\kern 1pt[Q_5^-,\bar \psi i\gamma _5I\psi ]\kern 1pt}
\right\rangle\quad\quad\quad\quad\quad\;\eqno(14)$$
$$\;\;\quad\quad\;\quad\quad\quad\simeq -{K(m) \over {\bar m}}\bar \theta {1
\over
{F_\pi }}\left\langle {\bar \psi \{{\textstyle{{\lambda ^8}
\over 2}},\kern 1pt\{{\textstyle{{\lambda ^+} \over 2}},
\kern 1pt\{{\textstyle{{\lambda ^-} \over 2}},I\}\psi }
\right\rangle={K(m) \over {\bar m}}\bar \theta
{1 \over {F_\pi ^3}}{2 \over {\sqrt 3}}C$$
where the pion decay constant $F_\p\approx 93 MeV$ and we have used
$[Q^a_5,F\tilde F]=0$. The broken generators of
$SU(3)_A$ corresponding to light pseudoscalars are given by
$$Q_5^a=\int {d^3x\psi ^{\dagger}\gamma _5{{\lambda ^a} \over 2}\psi (x)}
\quad\quad\;(a=1,2,\cdots ,8)\eqno{(15)}$$
where $\lambda^a$'s are Gell-Mann matrices and $\lambda^{\pm}=1/\sqrt{2}(
\lambda_1 \mp i\lambda_2)$. (14) has been first derived by
Crewther, Di Vicchia,
Veneziano and Witten (CDVW) \cite{8}. However, there have been doubts about the
calculation since it does not explicitly exhibit the
use of the topological non-triviality
of the $\q$-vacuum. More concretely, one may shift the strong CP phases from
$\q_{QFD}$ to $\q_{QCD}$ through chiral rotations and computes the amplitude,
as one does in (14),
$$\left\langle {\pi ^+\pi ^-\left| {{\cal L}_{(B)}^{CP}}
\right|\eta } \right\rangle=\bar \theta ({{-i}
\over {F_\pi }})\left\langle {\pi ^-\left| {[Q_5^-,F\tilde F]}
\right|\eta } \right\rangle\eqno{(16)}$$
which is zero if one imposes the canonical commutation relation by which
$Q^a_5$
commutes with gauge fields. This contradiction has triggered a  serious
doubt on whether or not the strong CP phases lead to any physical effects at
all
\cite{9}.

We believe that this concern is redundant. The vacuum alignment
equation (VAE) has
incorporated the non-perturbative features of QCD vacuum
into the game. Both ${\cal L}_{(A)}^{CP}$
and ${\cal L}_{(B)}^{CP}$ are solutions of the VAE and should,
if one does things
 correctly, result in the same conclusion. In basis (B)
the quark masses are real but
the condensates are complex. The vacuum does not respect CP
symmetry. In this case
even though ${\cal L}_{(B)}^{CP}$ does not contribute to the amplitude as
shown in (16), the CP conserving part of the lagrangian may do. Moreover, when
the quarks have non-degenerate masses (mass splitting), the condensates are of
the form
$$\left\langle {\bar \psi _L\psi _R} \right\rangle=-
{C \over 2}\beta \equiv -{C \over 2}(I+i\delta )\eqno{(17a)}$$
with
$$\beta =\left( {\matrix{{e^{i\beta _u}}&{}&{}\cr
{}&{e^{i\beta _d}}&{}\cr
{}&{}&\ddots \cr
}} \right)\;\;;\;\;\delta \simeq \left( {\matrix{{\beta _u}&{}&{}\cr
{}&{\beta _d}&{}\cr
{}&{}&\ddots \cr
}} \right)\eqno{(17b)}$$
if $\b_i$'s are small. Apparently (17a) is not invariant under $SU(L_f)_V$
transformations if $\b_i$'s are not all equal. In other words,
the vector charges defined as generators of $SU(L_f)_V$ do not annihilate
the vacuum completely or
$$Q^a\left| 0 \right\rangle\ne 0.\eqno{(18)}$$
Clearly, the  subgroup of $SU(L_f)_L \times SU(L_f)_R$ which leaves the vacuum
invariant
must satisfy
$$U_L^+\beta \kern 1ptU_R=\beta \quad\; or \;\quad U_R=\beta ^{-1}U_L
\beta \eqno{(19)}$$
where $U_L$ and $U_R$ are left and right unitary representations of
 $SU(L_f)$.
The broken generators, which excite the  glodstone bosons known as pions, are
those of the coset of the unbroken group. From (19) it is easy to understand
that the broken group is not $SU(L_f)_A$ any more but
to be rotated to $\b^{-1}SU(L_f)_A\b$ generated by $\d$.
The pion generators, denoted by $\tilde{Q}_5^a$, are thus
$$\tilde Q_5^a=\int {d^3x\{\psi ^{\dagger}\gamma _5{\textstyle
{{\lambda ^a} \over 2}}\psi (x)+}\psi ^{\dagger}[{\textstyle{{\lambda ^a}
\over 2}},i\delta ]\psi (x)\}+O(\delta ^2),\eqno{(20)}$$
i.\ e.\ , the pions are mixing of P-odd and P-even components.

Now that ${\cal L}_{(B)}^{CP}=\bq F \tilde{F}$ has no contribution to the
amplitude, we have
$${\cal A}(\eta \to \pi ^+\pi ^-)=\left\langle
{\pi ^+\pi ^-\left| {-\bar \psi \kern 1ptm\psi } \right|\eta } \right\rangle
\quad\quad\quad\quad\quad\quad\quad\quad\quad\quad\quad\quad\quad
\quad\quad\quad\quad\quad\quad\quad\quad$$
$$\quad\;\;\quad\quad=-({{-i} \over {F_\pi }})^3\left\langle
{[\tilde Q_5^8,[\tilde Q_5^+,[\tilde Q_5^-,\bar \psi
\kern 1ptm\psi ]} \right\rangle\quad\quad\quad\quad\quad\quad\quad\quad
\quad\quad\quad\quad\quad\eqno{(21)}$$
$$\quad\quad\;\quad\quad\quad=-{i \over {F_{\pi}^{3} }}\{\left\langle {\bar
\psi
[[{\textstyle{{\lambda ^8} \over 2}},i\delta ],\{{\textstyle{{\lambda ^+}
\over 2}},\{{\textstyle{{\lambda ^-} \over 2}},m\}\}]\psi }
\right\rangle+\left\langle
{\bar \psi \{{\textstyle{{\lambda ^8} \over 2}},
[[{\textstyle{{\lambda ^+} \over 2}}i\delta ],
\{{\textstyle{{\lambda ^-} \over 2}},m\}]\}\psi } \right\rangle$$
$$\quad\quad\quad\quad\;\quad+\left\langle {\bar \psi \{{\textstyle{{\lambda
^8}
\over 2}},\{{\textstyle{{\lambda ^+} \over 2}},[[{\textstyle{{\lambda ^-}
\over 2}},i\delta ],m]\}\}\psi } \right\rangle-\left\langle
{\bar \psi \gamma _5\{{\textstyle{{\lambda ^8} \over 2}},
\{{\textstyle{{\lambda ^+} \over 2}},\{{\textstyle{{\lambda ^-}
\over 2}},m\}\}\}\psi } \right\rangle\}$$
where $m$ is the diagonal mass matrix which is real in
this basis. The first three terms in parenthesis
come from the modification of the pion generators and the last term
reflects the complexity of the condensates which is absent in basis (A).
They are of the same order of $m$ and
$\bq$. Manipulations of these commutators yield, to $O(m;\bq)$
$$A(\eta \to \pi ^+\pi ^-)={{-i} \over {F_\pi ^3}}\{{i
\over {\sqrt 3}}(\beta _u-\beta _d)(-m_u\left\langle {\bar uu}
\right\rangle+m_d\left\langle {\bar dd} \right\rangle)\quad
\quad\quad\quad\quad\quad\quad$$
$$\quad\quad\quad\quad\quad\quad\quad-{1 \over {\sqrt 3}}(m_u+m_d)(\left\langle
{\bar u\gamma _5u}
\right\rangle+\left\langle {\bar d\gamma _5d} \right\rangle)\}={K(m) \over
{\bar m}}
\bar \theta {1 \over {F_\pi ^3}}{2 \over {\sqrt 3}}C.\eqno
{(22)}$$
In deriving the final step of (22) we have substituted in
$\langle\by^i\y^i\rangle=-C\cos \b_i\simeq -C$ and
 $\langle\by^ii\g_5\y^i\rangle=C\sin \b_i\simeq C\b_i$ and solution (12). We
 therefore confirm that CDVW's result is independent of  chiral frames.

We conclude that the study of the vacuum orientation of the dynamical
chiral symmetry breaking provides us an improved understanding of strong CP
violation. It has been shown that $\eta \rightarrow 2\pi$ decay occurs  if
$\bq$ is non-zero. More precise experiment measuremental on the decay rate is
encouraged to constrain  $\bq$ .

Huang would like to thank F. Gilman for arranging him a visit to SSC Lab when
this work was planned. A stimulation from S. Weinberg is gratefully
acknowledged.

\end{document}